# Magnetomechanical Effects in the Elastic Polymer Composites Containing Different Volume Fraction of Ferromagnetic Powder Particles.

A.A. Likhachev

*Institute for Metal Physics, National Academy of Sciences, 36 Vernadsky Str., 03680, Kiev, Ukraine,* e-mail: alexl@imp.kiev.ua

**Abstract**

In the present work a detailed thermodynamic consideration for the magnetic free energy of the composite material consisting of the ferromagnetic powder particles embedded into a polymer matrix is given. We estimate their magnetostatic interaction energy and it's dependence on the microscopic distribution of the magnetization and the magnetic field in the composite material. We also define the hydrostatic component of the mechanical force developed and the volume change effect caused by the magnetostatic interactions in such composites.

**Introduction**

In the last years, the study of heterogeneous materials consisting of magnetic micro- and nano-particles imbedded into a non-magnetic matrix has been increasing due to its importance for understanding micromagnetic interactions in these systems, as well as for possible applications [1, 2]. When the magnetic particles are magnetized and the matrix material is elastic, an elastic magnet is obtained. This material may exhibit elastomagnetic effects [3,4] and it can be used for sensors and actuators [5,6]. In order to have optimum performances for the mentioned applications, the material must have a high content of magnetic particles.

During these years several different materials were tested: magnetoelastic composite with the filling particles made of magnetostrictive, hard or soft ferromagnetic material [7,8]. Some possible applications in the airplane and car industries as actuators or anti-friction components [9]; heat-shrinkable elastic ferromagnets with the variable magnetic and conductive properties [10] were discussed. Those papers were dealing with a similar kind of materials and focusing on the theoretical and experimental correlation of the material elasticity with its magnetic behavior.

In particular, they analyzed a composite of particles uniformly dispersed inside the matrix material: a) the particles having an asymmetric shape, preferably with a main anisotropy axis; b) the particles that were soft ferromagnetic or small permanent magnets; c) the composites having an elastic behavior, due to the matrix properties, up to a relative deformation. In these conditions a strong coupling acts between magnetization axis and the main shape anisotropy axis of the particles. Therefore a change of the magnetizing field along an axis different from the easy magnetization one gives a rotation of the particles due to the mechanical torque, in order to align the magnetic moments with the applied field [11]. The macroscopic effect of these local rotations can be a deformation of the whole material. The inverse effect [12] consists in the change of magnetization axis due to a deformation of the elastic material. As an example, an elongation of the elastomagnetic material produces a rotation of each particle and a consequent rotation of its magnetic moment because it is strongly coupled with the particles geometry. This also gives a variation of the magnetization component along the elongation axis. In conclusion the inverse elastomagnetic effect can be used to have a strain sensor detecting deformation by means of the induced magnetization changes at constant temperature and magnetizing field [13-16].

A general target of the present work is to produce a detailed thermodynamic consideration for the magnetic free energy of the composite material consisting of the ferromagnetic powder particles embedded into a polymer matrix. We estimate their magnetostatic interaction energy and it's dependence on the microscopic distribution of the magnetization and the magnetic field in the composite. We also define the hydrostatic component of the mechanical force and the volume change effect caused by the magnetostatic interactions in such composites.

**Magnetic free energy, magnetic forces and strain effect in polymer composites containing ferromagnetic particles**

Generally, the magnetic forces can be produced in any ferromagnetic material when it magnetizes Everything depends on the fact if the material is deformable and if the magnetic free energy per unit volume $F^{mag}(\mathbf{h},\boldsymbol{\varepsilon})$ is dependent not only on the external magnetic field $\mathbf{h}$, but also on the strain of the material $\boldsymbol{\varepsilon}$. Here we have defined the magnetic free energy to be zero at $\mathbf{h}=0$. So, it represents only a magnetic part of the total free energy of the material $F(\mathbf{h},\boldsymbol{\varepsilon})=F(0,\boldsymbol{\varepsilon})+F^{mag}(\mathbf{h},\boldsymbol{\varepsilon})$. In such a case both the macroscopic magnetization of the material $\mathbf{m}(\mathbf{h},\boldsymbol{\varepsilon})$ and the field-induced magnetic forces $\boldsymbol{\sigma}^{mag}(\mathbf{h},\boldsymbol{\varepsilon})$ can be represented on the basis of the general thermodynamic relationships as follows:

$$\mathbf{m}^{mag}(\mathbf{h},\boldsymbol{\varepsilon})=-\left(\frac{\partial}{\partial \mathbf{h}}F^{mag}(\mathbf{h},\boldsymbol{\varepsilon})\right)_{\boldsymbol{\varepsilon}} \qquad \boldsymbol{\sigma}^{mag}(\mathbf{h},\boldsymbol{\varepsilon})=-\left(\frac{\partial}{\partial \boldsymbol{\varepsilon}}F^{mag}(\mathbf{h},\boldsymbol{\varepsilon})\right)_{\mathbf{h}} \qquad (1)$$

In absence of the magnetic field any deformable material can be strained elastically or unelastically applying the external mechanical load $\boldsymbol{\sigma}$. In that case the strain response of the material can be represented by its zero-field stress-strain relationship $\boldsymbol{\sigma}=\boldsymbol{\sigma}^0(\boldsymbol{\varepsilon})$. In general case, when both the mechanical and the magnetic forces are applied they work altogether simultaneously and a corresponding strain effect can be found from the following force balance equation:

$$\boldsymbol{\sigma}+\boldsymbol{\sigma}^{mag}(\mathbf{h},\boldsymbol{\varepsilon})=\boldsymbol{\sigma}^0(\boldsymbol{\varepsilon}) \qquad (2)$$

So, in two partial cases $\boldsymbol{\sigma}=0$ and $\mathbf{h}=0$, we obtain very similar relationships:

$$\boldsymbol{\sigma}^{mag}(\mathbf{h},\boldsymbol{\varepsilon})=\boldsymbol{\sigma}^0(\boldsymbol{\varepsilon}), \quad and \quad \boldsymbol{\sigma}=\boldsymbol{\sigma}^0(\boldsymbol{\varepsilon}) \qquad (3)$$

where, the first one defines implicitly the magnetic field induced strain (MFIS) effect in a particular ferromagnetic material, which is generally dependent both on its magnetic properties and also on the mechanical behavior of the material represented by its zero-field strain-stress relationship $\boldsymbol{\sigma}=\boldsymbol{\sigma}^0(\boldsymbol{\varepsilon})$. For instance, in the ordinary magnetostrictive materials it is given by a well-known linear and completely reversible Hook's law. In other recently discovered large MFIS systems like NiMnGa ferromagnetic shape-memory alloy the straining mechanism isn't elastic and based on the field-induced twinning occurring in these systems.

**Hydrostatic magnetic forces and volume change in polymer composites containing ferromagnetic particles.**

Here, we will apply the previously discussed idea to understand what the hydrostatic magnetic forces can be developed in systems consisting of the multiple ferromagnetic particles imbedded into some elastically soft polymer matrix and what the volume changes can be expected in similar materials.

For that, we introduce the total magnetic free energy $F$ of such a composite system of a volume $V$ and containing totally the volume $V_m$ of all ferromagnetic particles randomly distributed in the polymer matrix. We also assume that the saturation magnetization Ms of the bulk ferromagnetic material per its unit volume is known and that the saturation magnetization of each particular is the same.

In that case $F(V, V_m, h)$ can be represented in the following general form:

$$F(V,V_m,h) = VF_m\left(\frac{V_m}{V},h\right) = VF_m(z,h) \qquad (4)$$

Where, $F_m(z,h)$ is the magnetic free energy per unit volume of the composite, $h$ is the external magnetic field and $z = V_m/V$ is the volume fraction of the ferromagnetic material in the composite. This gives us the possibility to define the hydrostatic pressure developed in a magnetic field as follows:

$$P_m(z,h) = \frac{\partial}{\partial V}\left(VF_m\left(\frac{V_m}{V},h\right)\right)_h = F_m(z,h) - z\left(\frac{\partial}{\partial z}F_m(z,h)\right)_h \qquad (5)$$

We can also define the magnetization per unit volume of the composite material accordingly:

$$M(z,h) = -\frac{\partial}{\partial h}\left(F_m\left(\frac{V_m}{V},h\right)\right)_V = -\frac{\partial}{\partial h}(F_m(z,h))_z \qquad (6)$$

**Magnetostatic energy in polymer composites**

Generally, the total magnetic free energy of a composite material $F$ containing $N$ ferromagnetic particles magnetized to its full saturation consists of Zeeman's and the magnetostatic (demagnetizing) energy contributions. It can be represented as follows:

$$F = -\sum_{P=1}^{N} v_P\left(\mathbf{hm}_P + \frac{1}{2}\mathbf{h}_P\mathbf{m}_P\right). \qquad (7)$$

Here, $v_P$ are the particle volumes, $\mathbf{m}_P = M_s\mathbf{e}_h$ their local magnetizations all fully magnetized parallel to the external magnetic field $\mathbf{h}$ and $\mathbf{h}_P$ is the local demagnetizing field averaged over the particle volume.

The local demagnetizing field is produced by the surface magnetic charges induced both at the external surface of the composite material and also at the particle interface were normal magnetization components have jumps. So, it can be written in the following manner:

$$\mathbf{h}_P = -4\pi\mathbf{Dm} - 4\pi\mathbf{D}_P(\mathbf{m}_P - \mathbf{m}) \qquad (8)$$

where, $\mathbf{m}_P = M_s\mathbf{e}_h$, $\mathbf{m}$ is the macroscopic magnetization of the fully saturated composite material:

$$\mathbf{m} = \frac{1}{V}\sum_{P=1}^{N} v_P\mathbf{m}_P = zM_s\mathbf{e}_h; \qquad V_m = \sum_{P=1}^{N} v_P \qquad (9)$$

Here, $\mathbf{D}, \mathbf{D}_P$ are the demagnetizing matrices, representing both the composite material and the ferromagnetic particle dependent only on their shapes. So, finally, for the magnetic free energy per unit volume of the composite material we obtain:

$$F_m(z,h) = \frac{1}{2} 4\pi \left( \overline{D}_P z + (D - \overline{D}_P) z^2 \right) (M_s)^2 - z M_s h \qquad (10)$$

Here, $D, \overline{D}_P$ are the components of the demagnetizing matrices parallel to the external magnetic field applied. It is important, that $\overline{D}_P$ is defined as the average demagnetizing factor of the ferromagnetic particle system:

$$\overline{D}_P = \frac{1}{V_m} \sum_{P=1}^{N} v_P D_P; \qquad V_m = \sum_{P=1}^{N} v_P \qquad (11)$$

In particular, it means that if all the particles are randomly oriented then their average demagnetizing factor must be equal $\overline{D}_P = 1/3$ even if all of them are not really the spherical ones. In other words the ferromagnetic material distribution in the polymer matrix is statistically isotropic in this case.

Using Eqns. 5 and 10 one can also obtain the magnetic pressure:

$$P(z,h) = -\frac{1}{2} 4\pi (D - \overline{D}_P) z^2 (M_s)^2 \qquad (12)$$

The corresponding calculation results are represented in Fig.1.

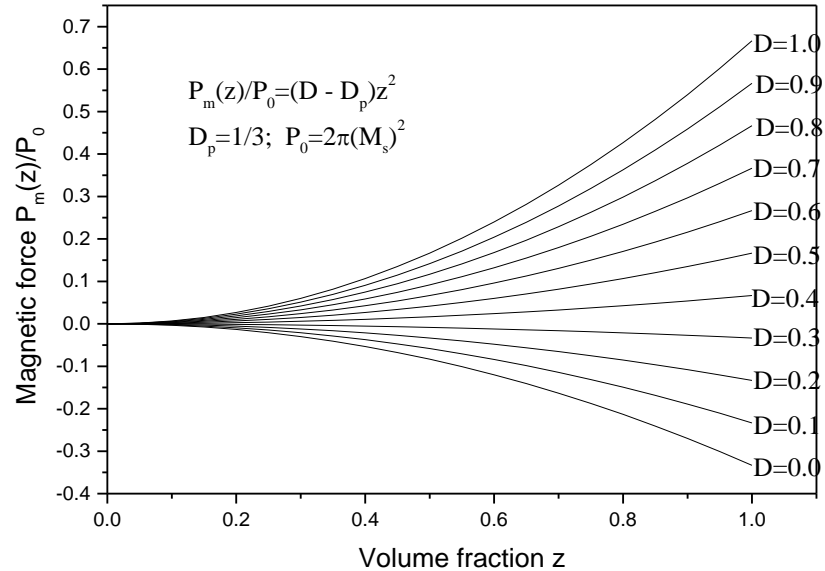

**Fig.1. Dependence of the magnetic pressure developed in a fully saturated composite containing the randomly oriented ferromagnetic particles at the different demagnetizing factors $D$ of the composite samples.**

As follows from these results the maximal magnetic force which is achieved in the completely saturated composite samples with the randomly oriented particles is strongly dependent on the volume fraction of the ferromagnetic material and the demagnetizing factor of the composite samples. It may have different signs as the demagnetizing factor changes within the range $0<D<1$. In case $D = 0$ (the sample is a long cylinder parallel to the magnetic field) the magnetic force will produce a compression effect. In the other case $D = 1$ (the sample is a thin plate aligned perpendicular to the magnetic field) the magnetic force will cause an extension. A general characteristic scale of the magnetic forces is defined by the material constant $P_0 = 2\pi(M_s)^2$, proportional to the saturation magnetization squared.

**Strain effect produced by magnetic forces**

As follows from Eq. 3., the relative volume change must be proportional to the magnetic force developed in a composite material and reverse proportional to its effective bulk elastic modulus $C(z)$.

$$\frac{\delta V}{V} = \varepsilon_{vol}(z,h) = K(z)P(z,h); \qquad K(z) = (C(z))^{-1} \qquad (13)$$

Here, $K(z)$ is obviously a bulk elastic compressibility of the composite material. As follows from the mechanical testing experiments shown in Fig.2 the elastic moduli of the polymer composites containing the rigid powder particles are strongly dependent on its volume fraction.

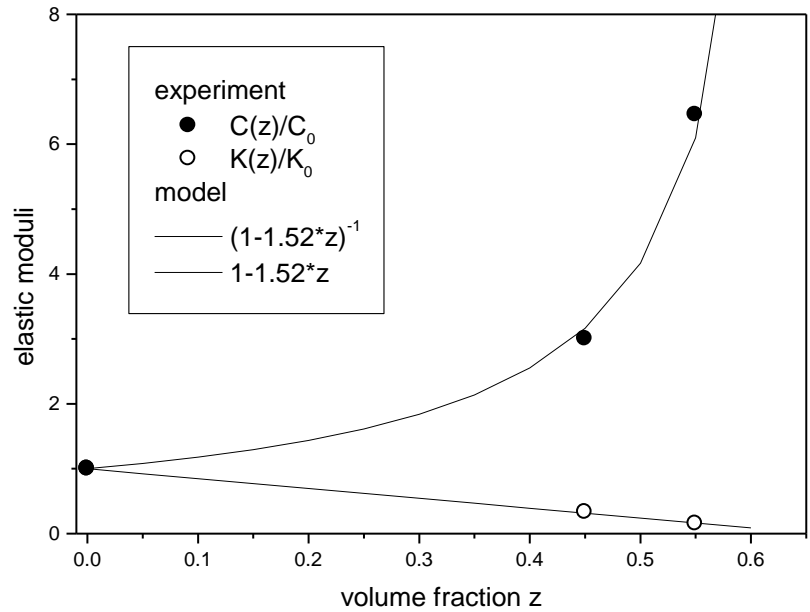

**Fig.2. Fraction dependence of bulk elastic modulus and elastic compressibility.**

These results indicate that the bulk elastic modulus is sharply goes up as the volume fraction of rigid particles increases so as the compressibility decreases approximately linearly. So, for some practical reasons we can use a following simple linear model for the elastic compressibility:

$$K(z) = K_0(1-z) \qquad (14)$$

It means that only the compressibility of polymer matrix $K_0$ gives a contribution into $K(z)$ proportional to its volume fraction $(1-z)$. So, finally we obtain the field induced volume effect:

$$\varepsilon_{vol}(z,h) = -2\pi K_0 (M_s)^2 (D - \overline{D}_P) z^2 (1-z) \qquad (15)$$

It is strongly dependent both on the sample and ferromagnetic particle shapes represented by the difference between their demagnetizing factors. It's also strongly dependent on the volume fraction of ferromagnetic material. The general scale of the field induced effects is defined by the material constant proportional to the elastic compressibility of the polymer and squared saturation magnetization of bulk ferromagnetic material. The fraction dependent factor has a maximum at the volume fraction at about $z = 2/3$. These final results are shown in Fig. 3.

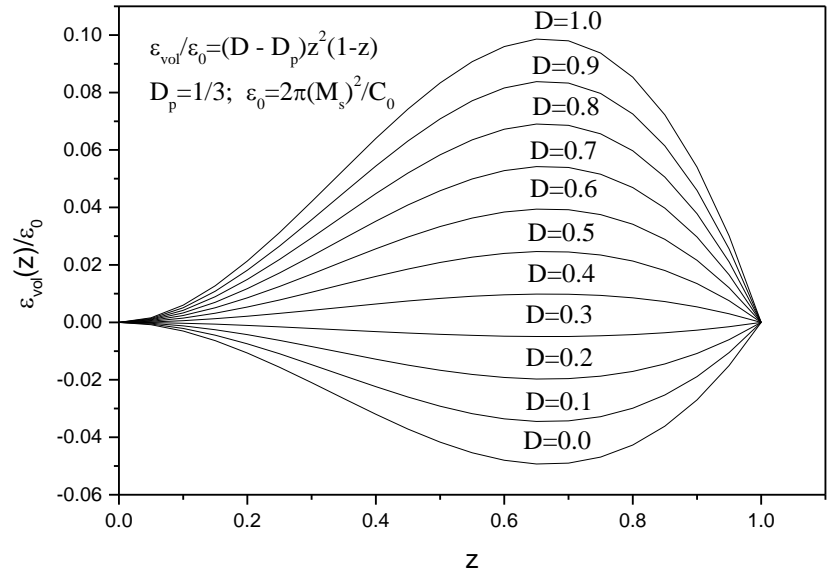

**Fig.3. Elastomagnetic volume change effect and its dependence on the volume fraction of ferromagnetic powder and demagnetizing factor of the composite sample.**

## Conclusions

The magnetic free energy of the composite material consists of the magnetostatic energy of separate ferromagnetic powder particles, their magnetostatic interaction energy and also of the Zeeman's energy. It is strongly dependent on the volume fraction of magnetic powder and the magnetic field applied.

Both the macroscopic magnetization and the hydrostatic magnetic forces can be obtained from the magnetic free energy as functions of the volume fraction and the magnetic field applied by using the general differential thermodynamic relationships.

The micromagnetic model of the polymer composite consisting of the ferromagnetic powders and elastically soft polymer matrix developed here has given the expression for the magnetostatic energy and the magnetic forces as functions of the volume fraction, magnetic field and the macroscopic demagnetizing factors of the composite samples. It has given a possibility to estimate the maximal field induced strain effect and its dependence on the volume fraction of the ferromagnetic material.

The experimental study of the elastic moduli of the polymer composites containing the different volume fraction of powder particles has shown it's strong dependence on the volume fraction of powder. One can use these results to estimate the fraction dependence of the field induced strain effect, it's dependence on the volume fraction of the ferromagnetic material and it's maximal value.

Finally, we have concluded that the field induced volume effect is strongly dependent both on the sample and ferromagnetic particle shapes represented by the difference between their demagnetizing factors. It's also strongly dependent on the volume fraction of ferromagnetic material. The general scale of the field induced effects is defined by the material constant proportional to the elastic compressibility of the polymer and squared saturation magnetization of bulk ferromagnetic material. The fraction dependent factor has a maximum at the volume fraction at about *z = 2/3*.

**Acknowledgements.** This work has been supported by Science Technology Center of Ukraine, Proj. No. 5522.

**References**

1. A.A. Novakova, V.Yu. Lanchinskaya, A.V. Volkov, T.S. Gendler, T.Yu. Kiseleva, M.A. Moskvina, S.B. Zezin. J. Magn. Magn. Mater. 258–259 (2003) 354.
2. H.M. Yin, L.Z. Sun, J.S. Chen, Mech. Mater. 34 (2002) 505.
3. L. Lanotte, G. Ausanio, V. Iannotti, C. Luponio Jr., Appl. Phys. A 77 (2003) 953.
4. L. Lanotte, G. Ausanio, C. Hison, V. Iannotti, C. Luponio, Sensors and Actuators A 106 (2003) 56.
5. G. Ausanio, C. Hison, V. Iannotti, C. Luponio, L. Lanotte, J. Magn. Magn. Mater. 272–276 (2004) 2069.
6. V. Iannotti, C. Hison, L. Lanotte, C. Luponio, G. Ausanio, C. Luponio, A. D'Agostino, R. Germano, Int. J. Appl. Electromagn. Mech. 19 (2004) 395.
7. T. A. Duenas, G. P. Carman, J. Appl. Phys. **87**, 4696 (2000).
8. S. Bednarek, Chinese journal of Physics **38**, 169 (2000).
9. M. Lokander, B. Stenberg, Polymer Testing **22**, 245 (2003).
10. S. Bednarek, Mat. Sci. Eng. B **77**, 120 (2000).
11. L. Lanotte, G. Ausanio, C. Hison, V. Iannotti, C. Luponio, Sensors and Actuators, **A 106**, 56 (2003).
12. L. Lanotte, G. Ausanio,V. Iannotti, G. Pepe, G. Carotenuto, P. Netti, L. Nicolais,
13. P. R. **B63**, 054438 (2001).
14. L. Lanotte, G. Ausanio, V. Iannotti, C. Luponio Jr, Appl. Phys A **77**, 953 (2003).
15. V. Iannotti, C. Hison, L. Lanotte, G. Ausanio, C. Luponio, A. D'Agostino, R. Germano, Proceeding ISEM2003, 246 (2003).
16. R. Zallen, The Physics of Amorphous Solids ed. Wiley, N.Y. (1983) Chapter 5, pp. 223.